\documentclass[11pt]{article}
\usepackage{epsfig}
\usepackage{amssymb}
\usepackage{amsmath}

\newcommand{\ee}{\end{equation}}
\newcommand{\eea}{\end{eqnarray}}
\newcommand{\be}{\begin{equation}}
\newcommand{\bea}{\begin{eqnarray}}

\begin{document}

\title{\LARGE \bf Spherical Non-Abelian Solutions in Conformal Gravity}
  \author{
  \large  Y. Brihaye$^a$ $\:${\em and}$\:$
  Y. Verbin$^b$ \thanks{Electronic addresses: brihaye@umh.ac.be; verbin@openu.ac.il } }
 \date{ }
   \maketitle 
    \centerline{$^a$ \em Physique Th\'eorique et Math\'ematiques, Universit\'e de Mons - UMONS,}
   \centerline{\em Place du Parc, B-7000  Mons, Belgique}
     \vskip 0.4cm
   \centerline{$^b$ \em Department of Natural Sciences, The Open University
   of Israel,}
   \centerline{\em Raanana 43107, Israel}

\maketitle
\thispagestyle{empty}
\begin{abstract}
We study static spherically-symmetric solutions of non-Abelian gauge theory coupled to Conformal Gravity. We find
solutions for the self-gravitating pure Yang-Mills case as well as monopole-like solutions of the Higgs system. The
former are localized enough to have finite mass and approach asymptotically the vacuum geometry of Conformal Gravity, 
while the latter do not decay fast enough to have analogous properties. 
\end{abstract}

\maketitle

\section{Introduction}\label{Introduction}
\setcounter{equation}{0}

Conformal Gravity \cite{Mannheim2006} (CG) was proposed as a possible alternative to Einstein gravity 
(``GR''), which may supply the proper framework for a solution to some of the most annoying problems of 
theoretical physics like those of the cosmological constant, the dark matter and the dark energy. 

It is therefore essential to investigate its predictions and consequences as further as 
possible. Here we choose to study static spherically-symmetric solutions of a non-Abelian gauge system coupled to
CG. This report may be regarded as a sequel to the previous one which dealt with field systems
with global symmetry and perfect fluid sources \cite{BrihayeVerbin}. 

The main ingredient of CG is the replacement of the Einstein-Hilbert action with 
the Weyl action based on 
the Weyl (or {\it conformal}) tensor $C_{\kappa\lambda\mu\nu}$ defined as the totally traceless 
part of the Riemann tensor (we use $R^{\kappa}_{\phantom{\kappa}\lambda\mu\nu}=
\partial_\nu \Gamma^{\kappa}_{\lambda\mu}-\partial_\mu \Gamma^{\kappa}_{\lambda\nu}+...$):
\begin{eqnarray}
C_{\kappa\lambda\mu\nu}=R_{\kappa\lambda\mu\nu}-
\frac{1}{2}(g_{\kappa\mu}R_{\lambda\nu}-g_{\kappa\nu}R_{\lambda\mu}+
g_{\lambda\nu}R_{\kappa\mu}-g_{\lambda\mu}R_{\kappa\nu})+
\frac{R}{6}(g_{\kappa\mu}g_{\lambda\nu}-g_{\kappa\nu}g_{\lambda\mu})
\label{WeylTensor},
\end{eqnarray}
so the gravitational Lagrangian is
\begin{equation}
{\cal L}_{g}= -\frac{1}{2\alpha}C_{\kappa\lambda\mu\nu}C^{\kappa\lambda\mu\nu} 
\label{GravL}
\end{equation}
where $\alpha$ is a dimensionless parameter. The gravitational field equations are 
formally similar to Einstein equations where the source is the energy-momentum tensor $T_{\mu\nu}$
and in the left-hand-side Bach tensor $W_{\mu\nu}$ replaces the Einstein tensor:
\begin{equation}
W_{\mu\nu} =  \frac{\alpha}{2} T_{\mu\nu} 
\label{GravFieldEq}
\end{equation}
Bach tensor is defined by:
\begin{eqnarray}
W_{\mu\nu}=\frac{1}{3}\nabla_\mu\nabla_\nu R-\nabla_\lambda\nabla^\lambda R_{\mu\nu}
+\frac{1}{6} (R^2+\nabla_\lambda\nabla^\lambda R-3R_{\kappa\lambda}R^{\kappa\lambda})g_{\mu\nu}+
2R^{\kappa\lambda}R_{\mu\kappa\nu\lambda}-\frac{2}{3}RR_{\mu\nu}
\label{BachTensor}
\end{eqnarray}
Since Bach tensor is traceless, CG can accommodate only sources with $T^\mu_\mu=0$. The Yang-Mills (YM) 
and Yang-Mills-Higgs (YMH) systems that will be considered here are of course compatible with this condition.

The general spherically-symmetric line-element may be written in terms of a single metric function by exploiting 
the conformal symmetry\cite{Mannheim2006}:
\begin{equation}
ds^2= B (r)dt^2 - dr^2/B (r) - r^2 ( d\theta^2+\sin^2 \theta d\varphi^2) 
\label{lineelsph}.
\end{equation}
The non-vanishing components of Ricci tensor and the Ricci scalar are then
\begin{eqnarray}
R^0_0 = R^r_r = -\frac{B''}{2}-\frac{B'}{r} \,\,\,\,\,\,; \,\,\,\,\,\, 
R^\theta_\theta = R^\varphi_\varphi = \frac{1-B}{r^2}-\frac{B'}{r} \,\,\,\,\,\,; \,\,\,\,\,\, 
R =\frac{2(1-B)}{r^2} -\frac{4B'}{r}-B''
\label{SphRicci}
\end{eqnarray}
and those of Bach tensor
\begin{eqnarray} \nonumber
W^0_0 = -\frac{1}{3r^4}+ B^2\left[\frac{1}{3r^4}+\frac{1}{3r^2}\left( \frac{B''}{B}+\left( \frac{B'}{B} \right)^2 -
\frac{2}{r}\frac{B'}{B} \right)-\frac{1}{3r}\frac{B'B''}{B^2}+
\frac{1}{12}\left( \frac{B''}{B} \right)^2 \right. \\ \left.
-\frac{1}{6}\frac{B'B'''}{B^2}-\frac{1}{r}\frac{B'''}{B}-\frac{1}{3}\frac{B''''}{B}\right]\\\nonumber
W^r_r =  -\frac{1}{3r^4}+ B^2\left[\frac{1}{3r^4}+\frac{1}{3r^2}\left( \frac{B''}{B}+\left( \frac{B'}{B} \right)^2 -
\frac{2}{r}\frac{B'}{B} \right)-\frac{1}{3r}\frac{B'B''}{B^2}+
\frac{1}{12}\left( \frac{B''}{B} \right)^2 \right. \\ \left.
-\frac{1}{6}\frac{B'B'''}{B^2}+\frac{1}{3r}\frac{B'''}{B}\right]\\\nonumber
W^\theta_\theta = W^\varphi_\varphi = \frac{1}{3r^4}
-B^2\left[\frac{1}{3r^4}+\frac{1}{3r^2}\left( \frac{B''}{B}+
\left( \frac{B'}{B} \right)^2 -\frac{2}{r}\frac{B'}{B} \right)
-\frac{1}{3r}\frac{B'B''}{B^2}+\frac{1}{12}\left( \frac{B''}{B} \right)^2 \right. \\ \left.
-\frac{1}{6}\frac{B'B'''}{B^2}-\frac{1}{3r}\frac{B'''}{B}-\frac{1}{6}\frac{B''''}{B}\right] 
\label{SphBach}
\end{eqnarray}
A useful property of these components is:
\begin{equation}
W^0_0-W^r_r=-\frac{B(rB)''''}{3r}\,\,\,\,\,\,; \,\,\,\,\,\, W^r_r+W^\theta_\theta=\frac{B(rB)''''}{6r}.
\label{W00WrrComb}
\end{equation}

In the following we will model the source by a static spherically symmetric  matter distribution, that is a 
(traceless) energy-momentum tensor of the form ${T}^{\mu}_{\nu}=diag(T^0_0,T^r_r,T^\theta_\theta,T^\theta_\theta)$. 
The ``inertial mass'' of the matter fields will be as usual:
\begin{equation}
M_I =  4\pi \int_0^\infty  dr r^2 T^0_0 (r) 
\label{IMassBS}
\end{equation}
Thanks to (\ref{W00WrrComb}) the gravitational field equations  
(\ref{GravFieldEq}) reduce to a single very simple equation:
\begin{equation}
\frac{(rB)''''}{r}= -\frac{3\alpha}{2B}(T^0_0-T^r_r)
\label{SphGravFieldEqPF}
\end{equation}
which has a similar structure to the fourth order ``Poisson equation''
\begin{equation}
\nabla^2 \nabla^2 u = -h  
\label{4thOrder}
\end{equation}
where $h(\textbf{r})$ is the source term. In the spherically symmetric case $\nabla^2 \nabla^2 u = (ru)''''/r$ and outside
a spherical source (or in vacuum) $u(r)$ is given by $u(r)=c_0 + c_1 r + c_2 /r + \kappa r^2$. The parameters  
are related to the source (assumed to extend within $r\leq a$) by
\begin{equation}
c_1 = \frac{1}{2} \int_0^a r^2 h(r) dr \,\,\,\,\,\,, \,\,\,\,\,\, c_2= \frac{1}{6} \int_0^a r^4 h(r) dr 
\label{SourceMoments}
\end{equation}
while $\kappa$ is free. Note that the volume integral of the matter 
density (i.e. of $h(r)$) turns up as the coefficient of the \textit{linear} term in the potential
rather than the $1/r$ one. It is related to the fact that in this theory the potential of a point particle is
linear in accord with the behavior of the Green function.

Similarly, in CG the general solution around a localized spherically-symmetric source is
\begin{equation}
B(r) = c_0 + c_1 r + c_2 /r + \kappa r^2 \,\,\,\,\,\,; \,\,\,\,\,\, c_0^2=1+3c_1 c_2
\label{ConfSch}
\end{equation}
where the additional relation between the coefficients comes from the $W^r_r = 0$ equation which is of a third order. 
We can easily express the two parameters of the exterior solutions by:
\begin{equation}
c_1 =  \frac{3\alpha}{4} \int_0^\infty dr r^2 (T^0_0 (r)- T^r_r (r))/B(r) 
\label{c1}
\end{equation}
\begin{equation}
c_2 =  \frac{\alpha}{4} \int_0^\infty dr r^4 (T^0_0 (r)- T^r_r (r))/B(r)
\label{c2}
\end{equation}
while $\kappa$ is still not fixed by the source. However, in this framework $\kappa$ may be considered as a 
cosmological constant such that $R=4\Lambda=-12\kappa$. We notice that taking $\alpha >0$ corresponds to gravitational 
attraction for ``normal'' matter with positive energy density and positive pressure.  

In the absence of a cosmological constant, the gravitational potential is asymptotically linear which enables one to explain 
the galactic rotation curves within this context \cite{MannKaz1989,Mannheim2006}. 

Among all the higher order gravitational theories \cite{Schmidt2006,Fabbri2008}, CG is unique in the sense that
it is based on an additional symmetry principle. The conformal symmetry imposes severe limitations 
on the allowed matter sources. When matter is described in terms of a Lagrangian, it is very much
constrained, but the Abelian and non-Abelian ($n$ generators $T^a$) Higgs models are essentially 
still consistent with the conformal symmetry provided the scalar field ``mass term'' is replaced 
with the appropriate ``conformal coupling'' term which introduces a non minimal coupling to the Ricci scalar 
$R$. The matter Lagrangian which we will use here is therefore
\begin{equation}
{\cal L}_{m}= \frac{1}{2}(D_\mu \Phi)^\dagger(D^\mu
\Phi)-\frac{1}{12}R|\Phi|^2 -\frac{\lambda}{4}|\Phi|^4 
-\frac{1}{4}F^a_{\mu\nu}F^{a \mu\nu} \label{matterL},
\end{equation}
where $D_\mu = \nabla_\mu - ieA^a_\mu T^a$ and $F^a_{\mu\nu}$ are the $n$ components of the Lie algebra-valued 
 field strength $F^a_{\mu\nu} T^a$. The resulting field equations are
\begin{eqnarray}
D_\mu D^\mu \Phi + \lambda |\Phi|^2 \Phi + \frac{R}{6} \Phi &=& 0 \label{FieldEqsScalar}\\
D_\mu F^{a \mu\nu} =-\frac{ie}{2}[\Phi^\dagger T^a (D^\nu \Phi)- (D^\nu \Phi)^\dagger T^a \Phi ]&=& J^{a\nu }
\label{FieldEqsVector}.
\end{eqnarray}

The gravitational field equations are (\ref{GravFieldEq}) with
\begin{equation}
{T}_{\mu\nu} = { T}_{\mu\nu}^{(minimal)}+{1\over 6}
\left (g_{\mu \nu} \nabla ^{\lambda }\nabla_
{\lambda } |\Phi|^2 - \nabla _{\mu }\nabla_{\nu } |\Phi|^2  - 
{G}_{\mu \nu} |\Phi|^2 \right)
\label{confTmn}
\end{equation}
${T}_{\mu\nu}^{(minimal)}$ being the ordinary (``minimal'') energy-momentum tensor of the 
Higgs model and ${G}_{\mu \nu}$ is the Einstein tensor.

Now we assume spherically-symmetric fields in the simplest non-trivial case of SU(2), namely 
$(T^a)^{bc} = -i\varepsilon^{abc}$ and $\Phi$ and $A_\mu$ take the 
``hedgehog'' and monopole forms which may be written most simply in terms of the spherical unit vectors in 3-space,
$\textbf{e}_r$, $\textbf{e}_\theta$ and $\textbf{e}_\varphi$:
\begin{eqnarray}\nonumber
\Phi^a &=& f(r)\textbf{e}^a_r\\
A^a_\mu dx^\mu &=& \frac{1}{e} \left ( a(r)-1 \right ) \left(\textbf{e}^a_\varphi d\theta- 
\textbf{e}^a_\theta \sin\theta d\varphi\right)\\
\nonumber
\end{eqnarray}

The components of the energy-momentum tensor are (using the $\Phi$-equation (\ref{FieldEqsScalar}) and the monopole
parametrization above):
\begin{eqnarray} 
T^0_0 &=& \frac{1}{3}\epsilon_s + \frac{1}{3}\epsilon_{sv} -\frac{1}{3}u +\epsilon_{v1} + \epsilon_{v2}  - 
\frac{f^2}{6} \left( R^0_0 - \frac{R}{6} \right) + \frac{B'}{12}(f^2)' \label{SphT00}\\
T^r_r &=& -\epsilon_s + \epsilon_{sv} +u -\epsilon_{v1} + \epsilon_{v2} - \frac{f^2}{6} \left( R^0_0 - \frac{R}{2} \right)
-\frac{1}{12}\left(B'+\frac{4B}{r}\right)(f^2)' \label{SphTrr}\\ 
T^\theta_\theta = T^\varphi_\varphi &=& \frac{1}{3}\epsilon_s - \frac{2}{3}\epsilon_{sv} -\frac{1}{3}u - \epsilon_{v2}  - 
\frac{f^2}{6} \left( R^\theta_\theta - \frac{R}{6} \right) + \frac{B}{6r}(f^2)' \label{SphTtr}\\
\nonumber
\end{eqnarray}
where we use the following abbreviations
\begin{eqnarray}
\epsilon_s=\frac{1}{2}Bf'^2 \;\; , \;\;\;\;\; \epsilon_{sv}=\frac{a^2 f^2}{r^2} \;\; ,\;\;\;\;\; u=\frac{\lambda}{4} f^4 
\;\; , \;\;\;\;\; \epsilon_{v1}=\frac{B a'^2}{e^2 r^2} \;\; , \;\;\;\;\; \epsilon_{v2}=\frac{(1- a^2)^2}{2 e^2 r^4}
\end{eqnarray}
and the explicit expressions for the Ricci tensor and scalar should be obtained from eq. (\ref{SphRicci}).

The field equations for the scalar and vector fields are the following second order equations 
\begin{equation}
\frac{\left(r^2Bf'\right)'}{r^2}-
\left(\frac{2 a^2}{r^2} + \frac{R}{6}\right)f -\lambda f^3 =0 
\label{SphScFieldEq}
\end{equation}
where one should again use (\ref{SphRicci}) for $R$, and
\begin{equation}
\left( Ba'\right)'+\frac{(1- a^2)a}{r^2} - e^2 f^2 a =0 
\label{SphVecFieldEq}
\end{equation}

Since there is only one independent metric component, it is obvious that not all the field equations 
(\ref{GravFieldEq}) are independent. Actually there is only one independent equation and we may use the 
third order one
\begin{equation}
W^r_r - \frac{\alpha}{2} T^r_r = 0
\label{ThrdOrdr}.
\end{equation}
However, a much simpler form is again obtained by using 
(\ref{W00WrrComb}) giving therefore the following fourth order equation for the metric component $B(r)$:
\begin{equation}
\frac{(rB)''''}{r}= -\frac{\alpha}{B}\left[ 2\epsilon_s - \epsilon_{sv} -2u +3\epsilon_{v1}  - \frac{R}{12} f^2
+\frac{1}{4}\left(B'+\frac{2B}{r}\right)(f^2)'
\right]
\label{SphGravFieldEq}.
\end{equation}

Actually, we can rescale the variables $r$ and $f$ by an arbitrary length scale $\ell$ such that we get the 
dimensionless variables $x=er/\ell$ and $\ell f$. The coupling constants also rescale as $\alpha / e^2$ and 
$\lambda / e^2$, and in terms of these we obtain the same equations as above with just substituting $e=1$ 
and replacing $r$ by $x$ (or thinking of $r$ as dimensionless). Since there is no natural scale in the system 
due to the conformal invariance, we may use the typical length of the scalar curvature say, $\kappa=1/\ell^2$, 
or as we actually did (for convenience) $\kappa\ell^2 =0.1$.
Solutions with different values of $\kappa$ are related to each other by a simple scaling law.

\section{Pure Yang-Mills Solutions}
\setcounter{equation}{0}

Self-gravitating pure YM solutions in GR were discovered by Bartnik and McKinnon \cite{BMK}
for asymptotically flat space-time and generalized in \cite{bjoraker} for
the case of asymptotically anti-de Sitter (AdS) space-time. In this section we will present the analogous of the latter solutions 
in CG. We analyze the solutions with the two possibilities of positive and negative $\alpha$. Indeed, 
negative value of $\alpha$ is a ``wrong sign'' choice since it yields a repulsive linear potential of 
localized solutions. However, the attractive contribution ($\kappa r^2$) from the negative cosmological constant is dominant. 
Therefore we do not exclude this possibility. As in the GR case, gravity can balance in certain circumstances the 
gauge fields self-repulsion to allow for a globally regular solutions.

The relevant set of equations is obtained by substituting $f=0$ in (\ref{SphVecFieldEq}) and (\ref{SphGravFieldEq}), 
and they will have the following dimensionless form in the particular case under consideration~:
\be
\label{equa_part}
x (xB)'''' = - 3 \alpha a'^2   \ \ ,\ \ \ \ \ \ x^2 (Ba')' + (1-a^2) a = 0
\ee

We solve the field equations with the boundary conditions:
\begin{eqnarray}
B(0)=1 \ \ , \ \ B'(0)=0 \ \ , \ \ B'''(0) = 0 \ \ , \ \  B''(\infty) = 2\kappa  \ \ , \  \ a(0)=1 \ \ , 
\  \ \ a(\infty)=a_0
\label{BCBMK}
\end{eqnarray}
which are necessary for regular localized solutions with finite inertial mass as well as finite mass parameters of 
the fourth order gravity, $c_1$ and $c_2$. The constant $\kappa$ is positive in order for space time to be asymptotically
AdS. Implementing these conditions in  the field equations (\ref{equa_part}) leads after some algebraic 
manipulations to the following asymptotic behavior of the gauge field:
\begin{equation}
a(x) = a_0 + \frac{a_1}{x} + \frac{a_2}{x^2} + \dots 
\label{YMaAsymp}
\end{equation}
For analyzing the asymptotic behavior of the metric tensor we use the following parametrization
\begin{eqnarray} 
B(x)= \kappa x^2 + B_1 x + B_0 - \frac{2m(x)}{x}
    \label{YMBAsympPar}
\end{eqnarray} 
where we further define $m(x)=m(\infty)-Q(x)/2x$. 
This form is very useful in getting the general behavior of $B(x)$ and obtaining the following asymptotic expression
\begin{eqnarray} \nonumber
B(x)= \kappa x^2 
    +  \frac{a_0^3 - a_0 - 2 \kappa a_2}{a_1} x 
    +  \frac{-4 a_2 a_0^3 + 3 a_1^2 a_0^2 + 4 a_2 a_0 - a_1^3 + 8 \kappa a_2^2 - 6 a_1 a_3 \kappa}{2 a_1^2} + 
    \\ \nonumber
     \frac{(12 a_2^2 -9 a_1 a_3)a_0^3 - 6 a_2 a_1^2 a_0^2 + 3(a_1^4 - 4a_2^2+ 3 a_1 a_3)a_0 
              + 2 a_2 a_1^2 + 12\kappa (3 a_1 a_2 a_3 -2 a_2^3 - a_1^3 a_4 )}{3 a_1^3 } \frac{1}{x} \\
    - \frac{\alpha a_1^2}{8 } \frac{1}{x^2}
    - \frac{\alpha a_1 a_2}{10 } \frac{1}{x^3} 
    -\frac{\alpha (3 a_1 a_3 + 2 a_2^2)}{60} \frac{1}{x^4} +  \dots \hspace {7.6cm} 
    \label{YMBAsymp}
\end{eqnarray} 
with arbitrary parameters $\kappa, a_0, ... , a_4$. 
In fact, $\kappa$ and $a_0$ are fixed  by the boundary conditions. 

Whenever it exists, a solution corresponding to fixed $\alpha$, $a_0$ and $\kappa$, 
has definite values of the parameters $a_1, a_2, a_3, a_4$. 
They can be extracted from the numerical results described below. 

\begin{figure}[!b]
\centering
\leavevmode\epsfxsize=10.0cm
 \includegraphics[height=.30\textheight,width=.48\textwidth]{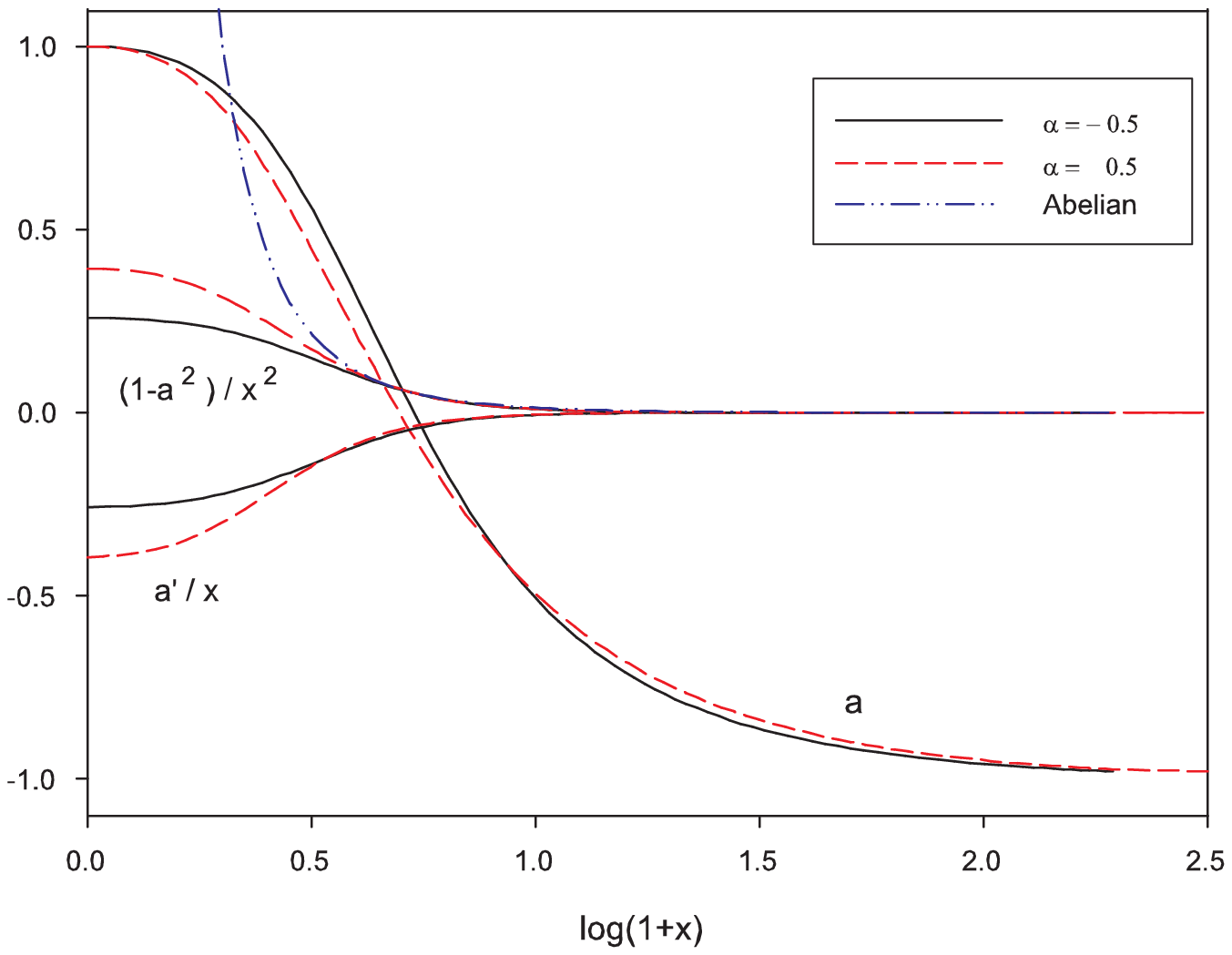}  
 \includegraphics[height=.30\textheight,width=.48\textwidth]{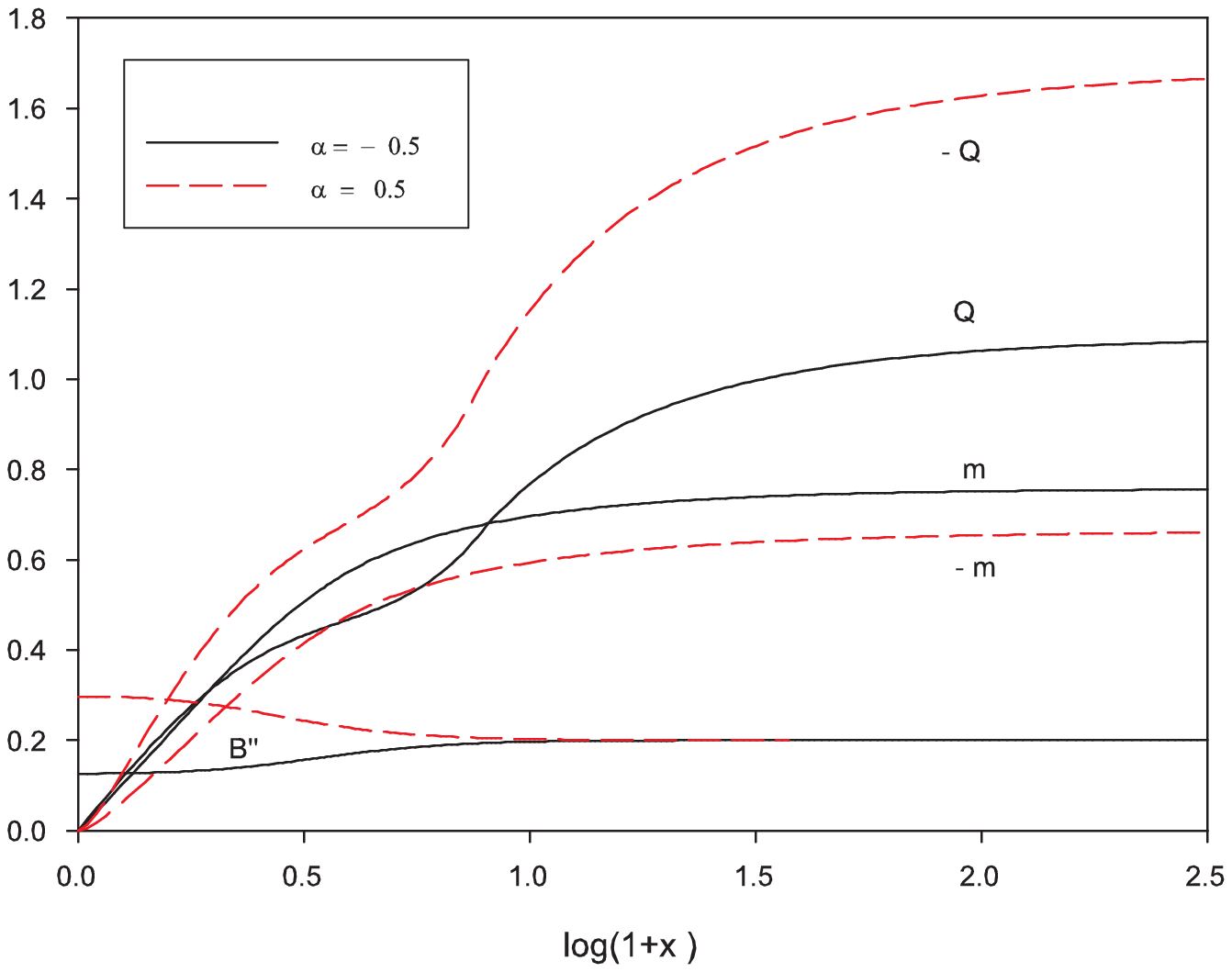}\\
 (a)\hskip 7.5cm (b)\\
 \caption{\label{Profiles2SignBMK+A} \small{Profiles of pure YM solutions with $\alpha=\pm 0.5$.
 The other parameters are $a_0=-1$ (or $Q_m=0$) and $\kappa=0.1$. 
(a) The gauge components and their corresponding magnetic fields. 
 Also added is the Abelian field strength.
(b) The metric functions: $B''$, $m(x)$ and $Q(x)$.}}
\end{figure}  

\begin{figure}[!t]
\centering
\leavevmode\epsfxsize=10.0cm
 \includegraphics[height=.30\textheight,width=.48\textwidth]{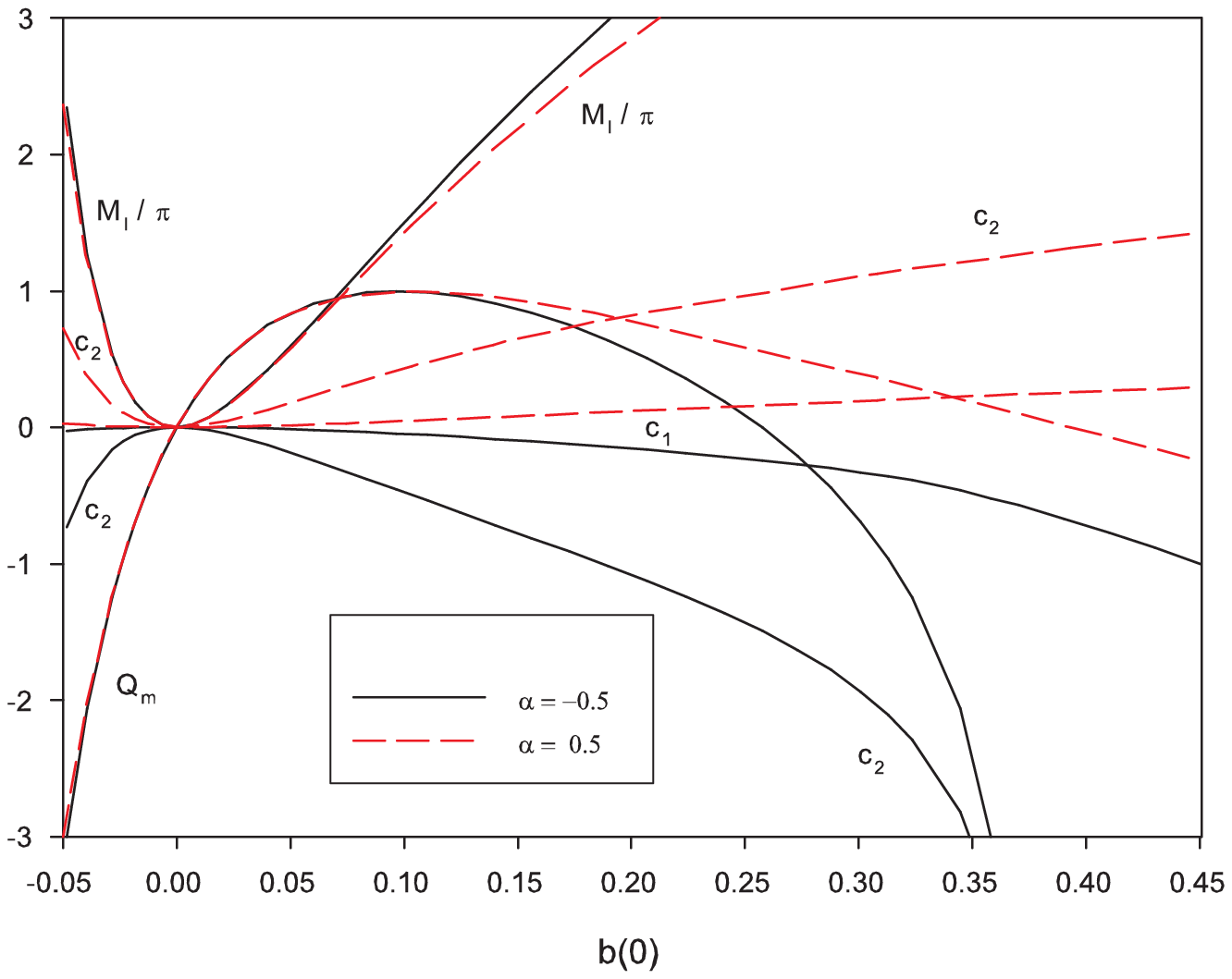}  
 \includegraphics[height=.30\textheight,width=.48\textwidth]{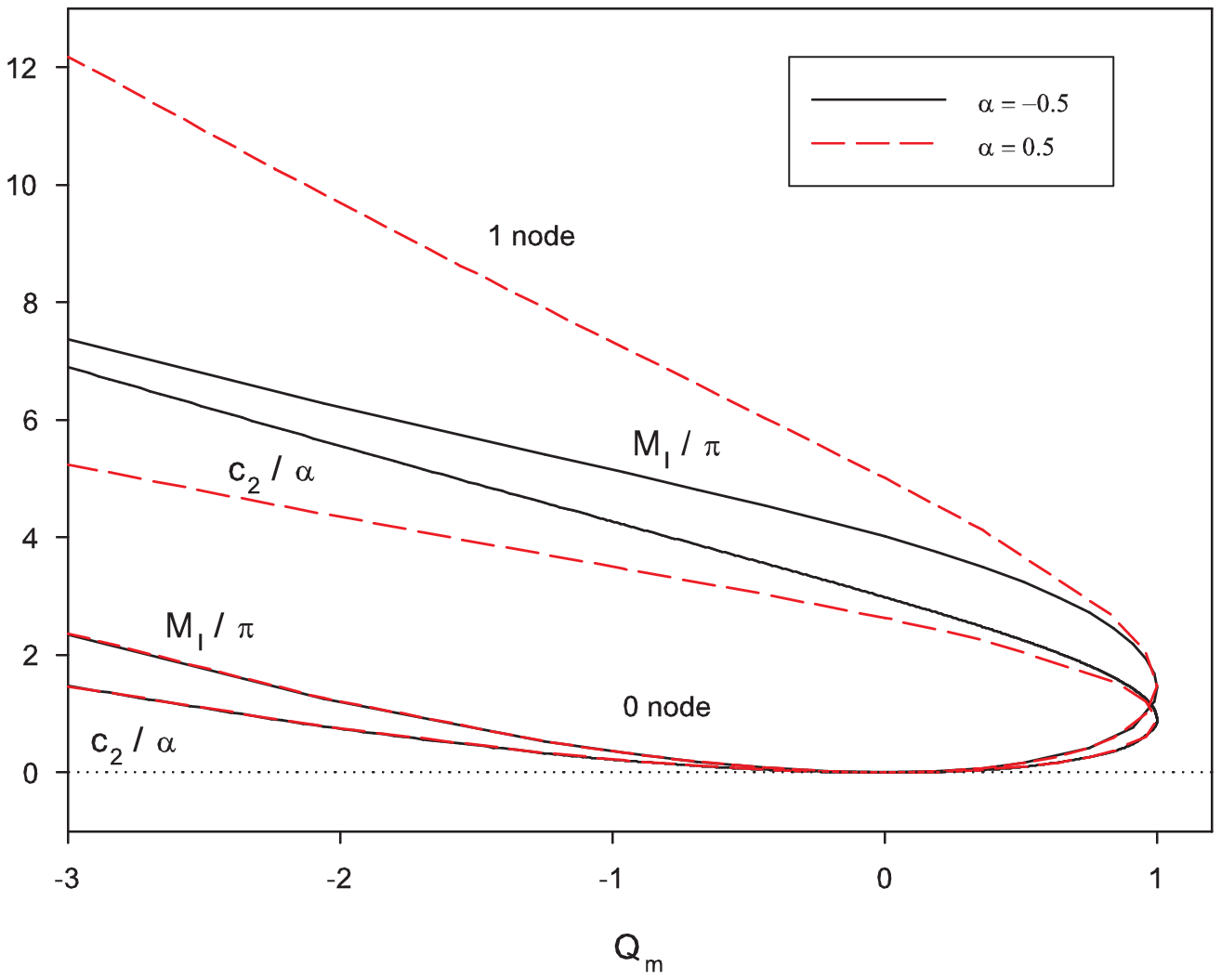}\\
 (a)\hskip 7.5cm (b)\\
 \caption{\label{data_bm_a1pm} \small{Several properties of pure YM solutions for two opposite values of 
 $\alpha$: $\alpha=\pm 0.5$.  $\kappa=0.1$ (a) Dependence on the central
magnetic field $b(0)$. Note that all lines cross at $b(0)=0$ which corresponds to the trivial solution with $a_0=1$.
(b) Dependence on the magnetic charge $Q_m$.}}
\end{figure}

Along with Ref. \cite{bjoraker} we find that the gauge function $a(x)$ can approach an arbitrary constant ($a_0$) at 
infinity, leading to  a continuous family of intrinsically different solutions with continuously varying magnetic charge
(or flux) $Q_m=1-a_0^2$. This contrasts the asymptotically flat case (in GR) where the field $a(x)$ should approach only the 
values $a(\infty) = \pm 1$ which correspond to vanishing magnetic charge. Actually, it seems that 
the analogous  asymptotically flat solutions in CG do not exist. 
  Our solutions comprise therefore a two-parameter ($\kappa$, $a_0$) family for any
given $\alpha$. The parameter $\kappa$ can be set to a fixed value by an appropriate scaling of the radial variable. 

It may be also of some interest to recall the analogous solutions of the Abelian case which are known in an explicit
form  \cite{mannheim_kazanas_2}. These solutions may carry both magnetic and electric charges 
denoted $p$ and $q$:
\begin{equation}
\frac{1}{2}F_{\mu\nu}dx^\mu\wedge dx^\nu =  \frac{q}{r^2} dt\wedge dr + p \sin\theta d\theta \wedge d\varphi 
\label{ConfRNgauge}
\end{equation}
and the Weyl equations lead to solutions of the form  (\ref{ConfSch}) for the field $B(r)$ where the relation between
the coefficients is modified by a contribution from the magnetic and electric charges:
\begin{equation}
B(r) = c_0 + c_1 r + c_2 /r + \kappa r^2 \,\,\,\,\,\,; \,\,\,\,\,\, c_0^2=1+3c_1 c_2 + \frac{3\alpha}{4}(p^2+q^2)
\label{ConfRN}
\end{equation}
\begin{figure}[!b]
\centering
\leavevmode\epsfxsize=8.5cm
\epsfbox{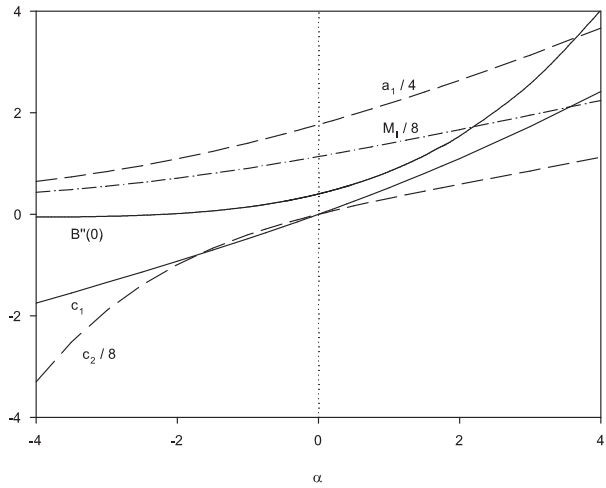}\\
\caption{\label{alpha_vary_BMK}  \small{$\alpha$-dependence of several properties of the pure YM solutions. 
The other parameters are $a_0=-1$ (or $Q_m=0$) and $\kappa=0.1$. Note that $M_I$ is positive for all $\alpha$, but 
the signs of both $c_1$ and $c_2$ are correlated with that of $\alpha$.}}
\end{figure}
  In absence of explicit solutions, we approached the above non-linear system of equations numerically and found that 
  solutions exist for generic values of the parameters $\alpha$ and $a_0$.
  
  Our results are illustrated by the three attached figures. The profiles of solutions with two opposite values of $\alpha$
  are presented in Fig. \ref{Profiles2SignBMK+A}. By inspection of the equations, it turns out that the
  sign of $\alpha$ is very much apparent from the behaviour of $B''(x)$; this is indeed what
  we observe on the figure. Furthermore, the figure shows that the magnetic field as well as $m(x)$ and $Q(x)$
  are also  sensitive to the sign. Note that the two magnetic components, the ``transverse''  $-a'/x$ (which
  we denote $b(x)$ for further use) and   the ``radial''  $(1-a^2)/x^2$ are very similar, but not identical as 
  shown by a closer inspection. Fig. \ref{Profiles2SignBMK+A} also  reveals that the fields reach their asymptotic values 
  rather gradually unlike the Bartnik-McKinnon solutions \cite{BMK} whose structure clearly split in three different regions 
  of space.

  The dependence of several parameters characterizing the solutions on the central magnetic field
   $b(0)$ and on the magnetic charge $Q_m$ is shown on Fig. \ref{data_bm_a1pm} -- again for two opposite $\alpha$'s. 
  We note the strong similarity between  our Fig. \ref{data_bm_a1pm}b and the corresponding plot of Ref. \cite{bjoraker}.
  The $Q_m(b(0))$ curves of Fig. \ref{data_bm_a1pm}a have a similar general structure to that of Ref. \cite{bjoraker},
  but an exact comparison cannot be made since our solutions are purely magnetic. 
  
 An additional plot (Fig. \ref{alpha_vary_BMK}), shows the dependence of the solutions on $\alpha$. This figure reveals the 
   non linear response   to $\alpha$ of the solutions  and exhibits in particular the asymmetry between positive and 
   negative $\alpha$.  
  
 A careful inspection of these three figures reveals that they exhibit a comforting agreement with each other. 
  For example, the two values of $b(0)$ in Fig. \ref{Profiles2SignBMK+A}a are equal to those obtained from the 
  points where $Q_m=0$  (which correspond to $a_{0}=-1$) in Fig. \ref{data_bm_a1pm}a. The values of $c_1$ 
  and $c_2$ at the same $b(0)$ values in Fig. \ref{data_bm_a1pm}a   are equal to those in Fig. \ref{alpha_vary_BMK}
   taken at $\alpha=\pm 0.5$.   
  
  These solutions fit nicely to the general discussion in the introduction about localized solutions
  in CG. The coefficient of the linear term in the asymptotic expansion (\ref{YMBAsymp}) is verified to be 
  equal to $c_1$   calculated directly from (\ref{c1}). Similarly, the coefficient $c_2$ of the $1/x$ term can also be 
  obtained both ways.  Note also that $M_I$ is positive for all $\alpha$, but the signs of both $c_1$ and $c_2$ are 
  correlated with that of $\alpha$.


It is natural to compare the solutions obtained in this section with the Abelian solutions mentioned above. 
These solutions possess both magnetic and electric charges.

The magnetic Abelian solution is embedded in our system as the very simple solution $a(x)=0$ (of Eq. (\ref{equa_part})) 
which has an inverse-square behavior of the field strength. The electric solution may be obtained just by duality. 
In both cases the Weyl equations lead to the gravitational field given by Eq. (\ref{ConfRN}).  
In particular, the expansion (\ref{YMBAsympPar}) is truncated to the $1/x$ term, the field $B(x)$ presents a horizon at some
 $x=x_h$, both the metric and the magnetic field are not defined at the origin. 

 In contrast, the non-Abelian solutions have non-trivial $a(x)$; the coupled equations (\ref{equa_part}) 
lead to the full  multipole series given above for both $a(x)$ and $B(x)$. 
Generic YM (magnetic) fields are regular at the origin, namely have  $a_2 \neq 0$ (as  confirmed 
numerically). The mass parameter $c_2$, is non-vanishing and was determined numerically 
(the dependence  on $\alpha$ is shown on Fig. \ref{alpha_vary_BMK}). The next correction (the $1/x^2$ term) is also
non-zero for generic solutions.

\section{Monopole-Like Solutions}
\setcounter{equation}{0}

Monopoles naturally emerge as topological defects in spontaneously broken theories and their non observation  
leads to some constraints on a large number of particle physics and cosmological models. Similarly, the degree
of physical relevance of CG depends on the results of an analogous study.  This leads us to examine if monopoles survive 
in this theory and, in the case they do, to study their qualitative properties.

Gravitating non-Abelian monopoles  (and their black holes counterparts) were studied by Ortiz \cite{Ortiz1992}, 
Lee et al. \cite{LNW1992} and Breitenlohner et al. \cite{BFM1992,BFM1995} both 
in asymptotically flat and AdS spaces.
 
In parallel with the previous section, we obtain here the counterpart of these solutions in CG.
In this case we  get solutions with asymptotically  broken symmetry if $R$ tends to a negative constant 
(that is $R\to-12\kappa$).  In fact, this system has been studied already by Edery et al. \cite{EderyEtAl2006,EderyEtAl2009}
but only for $\alpha=-1/2$. We go beyond these first results, addressing the domain of existence
of solution in the $(\lambda,\alpha)$ plane and studying some physical properties of the solutions; this needs in particular
a better understanding of the asymptotic behaviour of the solutions as is done below. 

We solve the field equations with the usual boundary conditions for regular and localized solutions:
\begin{eqnarray}
B(0)=1 \ \ , \ \ B'(0)=0 \ \ , \ \ B'''(0) = 0 \ \ , \ \  B''(\infty) = 2\kappa \\ \nonumber
f(0) = 0 \ \ , \ \  f(\infty)=f_\infty  \ \ , \ \ \ a(0)=1 \ \ , \  \ \ \ a(\infty)=0
\label{BCMon}.
\end{eqnarray}

The new characteristic here (with respect to the GR case) is that the vacuum expectation value of the Higgs field is not 
fixed by the Lagrangian, but by the asymptotic curvature namely $f_\infty = \sqrt{2\kappa/\lambda}$ 
(assuming as usual $\lambda > 0$). 

Performing a detailed asymptotic analysis of the solutions with the above boundary conditions at $x\to \infty$
leads to several possible types of behaviour. A combination of analytical and numerical considerations
reveals that only one of these possibilities matches with the regularity conditions at the origin.
The resulting asymptotic form is then~:

\begin{equation}
a(x)\sim a_1 / x^s + \cdots \ \ , \ \ s=\frac{1}{2}\left(1+\sqrt{1+\frac{8}{\lambda}}\right)
\label{aAsymp}
\end{equation}

\begin{equation}
f(x)\sim f_\infty+f_1 / x^p + \cdots \ \ , \ \ p=\frac{1}{2}\left(3-\sqrt{1-\frac{2\alpha}{3\lambda}}\right)
\label{fAsymp}
\end{equation}

\begin{equation}
B(x)\sim B_0+B_1 x^q + \kappa x^2 + \cdots \ \ , \ \ q=\frac{1}{2}\left(1+\sqrt{1-\frac{2\alpha}{3\lambda}}\right)
\label{BAsymp}
\end{equation}

\begin{figure}[!b]
\centering
\leavevmode\epsfxsize=10.0cm
 \includegraphics[height=.26\textheight,width=.48\textwidth]{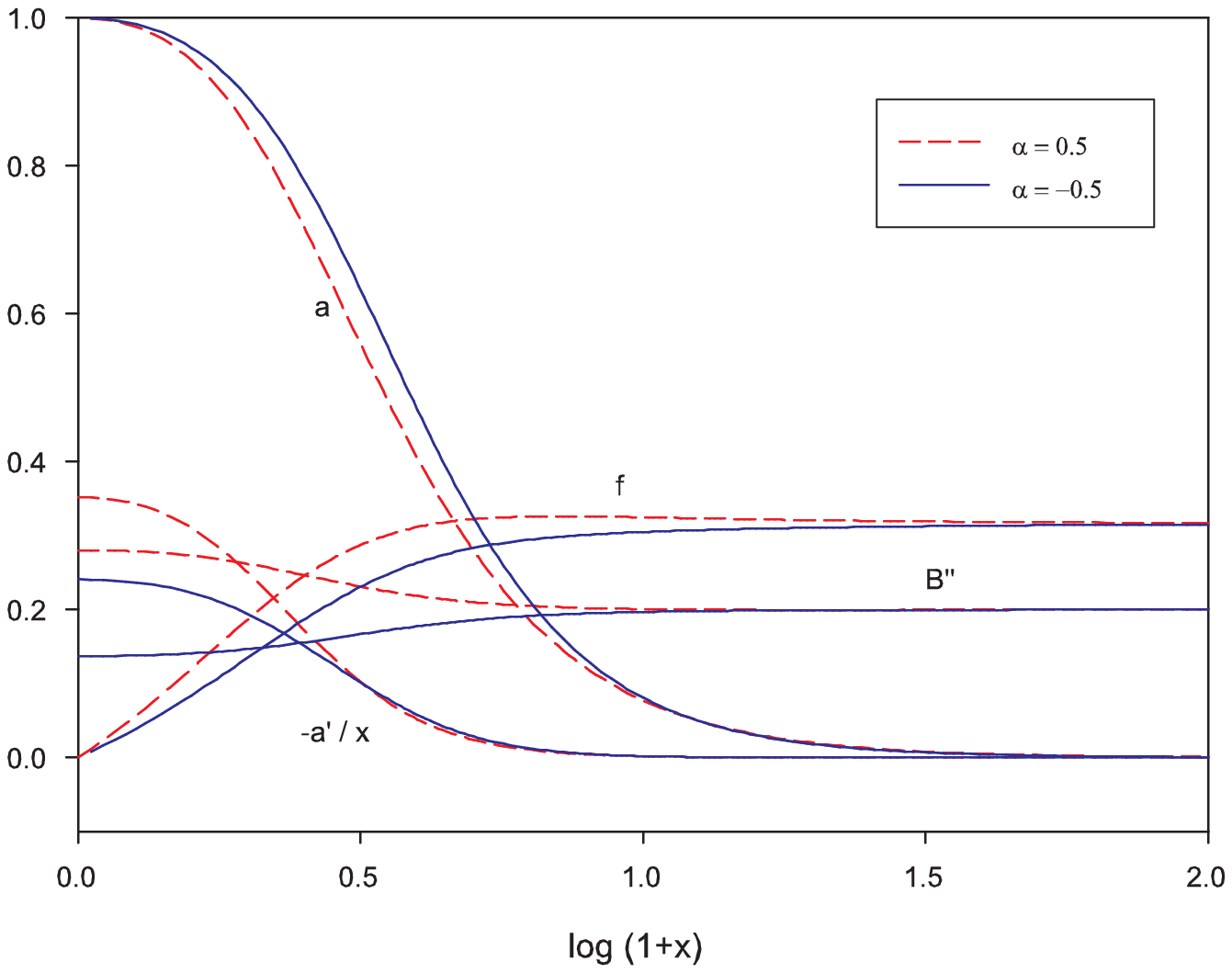}  
 \includegraphics[height=.26\textheight,width=.48\textwidth]{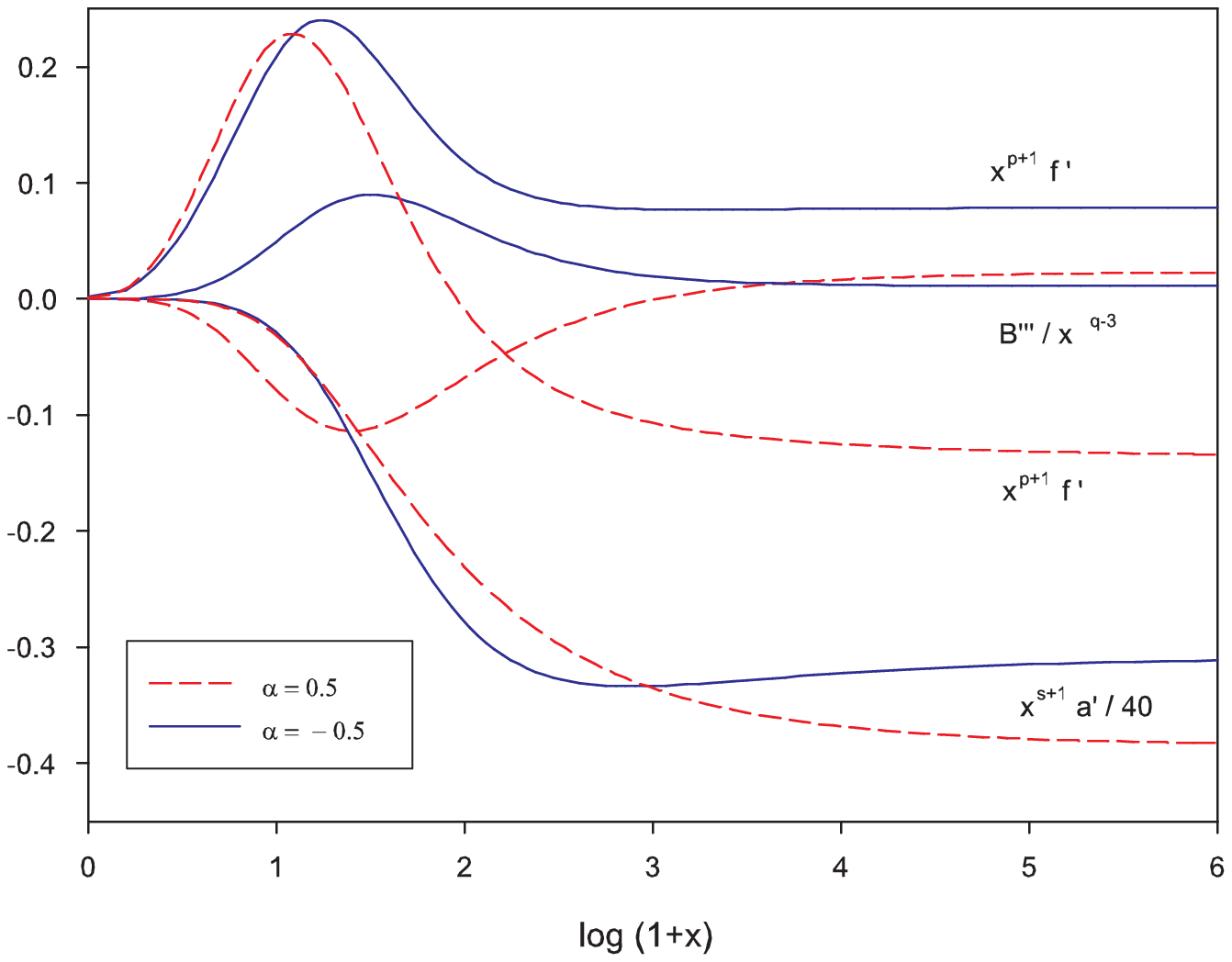}\\
 (a)\hskip 7.5cm (b)\\
\caption{\label{DetailesProfiles2Sign}  \small{Monopole-like solutions with two opposite values of $\alpha$: $\alpha=\pm 0.5$. 
The other parameters are $\lambda=1$ and $\kappa=0.1$.
(a) the profiles of the solutions; 
(b) demonstration of the asymptotic behavior of $f(x)$, $a(x)$ and $B(x)$. }}
\end{figure}

We note the presence of the parameter $\alpha$ in the exponents of the fields $B(x)$ and $f(x)$, but not in $a(x)$.
The additional term with respect to the usual quadratic term of AdS space time is therefore non-polynomial.
For $\alpha > 0$, the exponents in $B,f$ provide a natural upper limit in the
domain of existence of the solutions~:  $\alpha/\lambda < 3/2$.
A similar pattern is also present in the case of the GR monopole \cite{Ortiz1992,LNW1992,BFM1992,BFM1995}. 
Note also that the condition $p>0$ (or $q<2$) introduces a lower limit as well,  $\alpha/\lambda > -4$,
which is effective for $\alpha <0$. Both conditions can be also viewed as conditions on $\lambda$ for a fixed $\alpha$
since small enough $\lambda$ may violate them. We have limited our numerical investigation
to values of the parameters $\alpha,\lambda$ such that the field obeys the form above with a good accuracy.
For instance to $\alpha \in [-0.5,0.5]$, $\lambda \in [0.5,2]$. We expect however the solutions to exist for 
arbitrarily large values of $\lambda$ within the above-mentioned domain. 
Finally we notice that the limit $\lambda \to 0$ is singular even together with $\alpha \to 0$.  

The asymptotic behaviour above causes the energy density to decrease only as $T^0_0 \sim 1/x^p$ 
while the radial pressure decays more rapidly.
Therefore, the inertial mass (\ref{IMassBS}) and the mass parameter  $c_2$, eq. (\ref{c2})
clearly diverge; so does the coefficient of the linear term in the potential 
$c_1$, eq. (\ref{c1}) for $\alpha < 0$. For $\alpha >0$ the integral for $c_1$ seems to converge since then $p>1$,
 but simple integration of the 
field equation (\ref{SphGravFieldEq}) shows that it actually vanishes.
As a consequence, the solutions, although decaying to the vacuum as $x\to \infty$, are not localized enough to 
have a finite inertial mass. This can be seen from the asymptotic behavior of the solutions.

\begin{figure}[!t]
\centering
\leavevmode\epsfxsize=8.5cm
\epsfbox{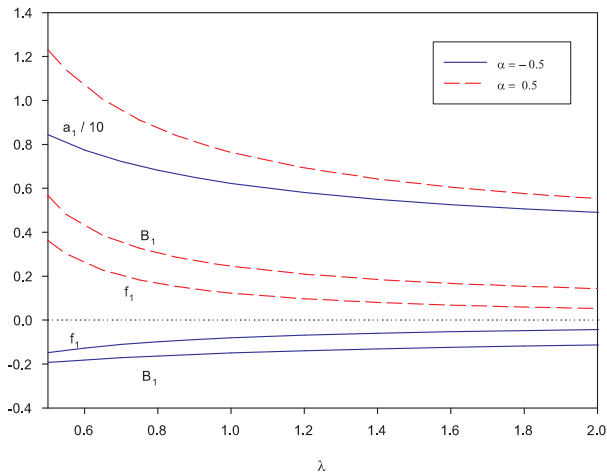}\\
\caption{\label{data_mono_bis} \small{$\lambda$-dependence of several properties of monopole-like solutions for two 
opposite values of $\alpha$: $\alpha=\pm 0.5$.  $\kappa=0.1$. Note the consistency with Fig. \ref{DetailesProfiles2Sign}; 
e.g. for $\alpha<0$, $f(x)$ increases towards its asymptotical value, so $f_1<0$, while for $\alpha>0$, $f(x)$ 
decreases towards its asymptotical value, so $f_1>0$.}}
\end{figure}

\begin{figure}[!b]
\centering
\leavevmode\epsfxsize=8.5cm
\epsfbox{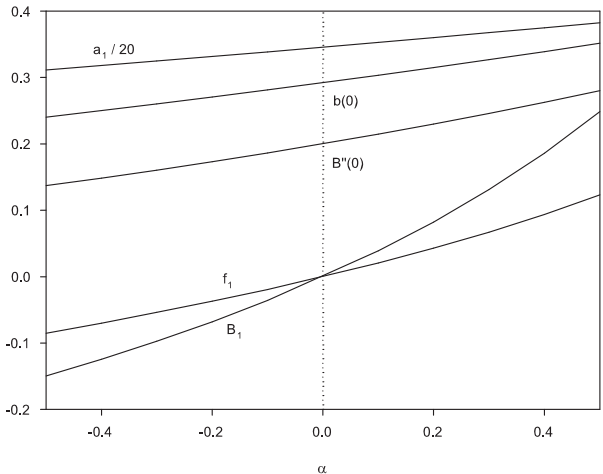}\\
\caption{\label{alpha_vary}  \small{$\alpha$-dependence of several properties of monopole-like solutions. We denote 
$b(x)=-a'(x)/x$. The other parameters are $\lambda=1$ and $\kappa=0.1$.}}
\end{figure}

We analyze the solutions with the two possibilities of positive and negative $\alpha$. 
In passing we can note  that, in the case $\alpha=0$,
gravity decouples from the gauge system. Since the corresponding 
vacuum geometry is AdS space, the scalar and vector fields then  
constitute a monopole solution in an AdS background \cite{LugoEtAl1,LugoEtAl2}.

We have constructed the numerical solutions and studied them for both signs of $\alpha$.
As mentioned already, negative $\alpha$  is a ``wrong sign'' choice since it yields a repulsive linear potential of 
localized solutions, but we do not exclude this possibility since a negative cosmological constant is induced by the 
spontaneous symmetry breaking and the attractive $\kappa r^2$ is dominant. The case 
$\alpha = -1/2$ is the one considered by Edery et al. \cite{EderyEtAl2006,EderyEtAl2009}. The relation with their 
parameter $k$ is $\alpha = -k/2$. (Actually, there is also a difference of a $\sqrt{2}$ factor 
between their $f$ and ours.)

The profiles of typical solutions are shown in Fig. \ref{DetailesProfiles2Sign} for $\alpha =\pm 0.5$.
We see a significant difference between the two signs of $\alpha$.
 For negative  $\alpha$, the field $f(x)$ approaches its asymptotic value in a monotonically increasing way 
 (accordingly, the  coefficient $f_1$ is  negative). For the opposite case 
   $f(x)$ crosses the value of $f_\infty$ and then approaches
this value from above. The coefficient $f_1$ is now positive. Similarly the behaviour of the function $B''(x)$
is affected by the sign. This function starts increasing from its minimal value at the origin for $\alpha<0$
and approaches monotonically its asymptotic value $2\kappa$; for $\alpha>0$, $B''(x)$ starts decreasing from
its maximal value and attains its asymptotic value with one oscillation which is not perceptible on the plot. 

Fig. \ref{DetailesProfiles2Sign}b demonstrates the asymptotic behavior given in eq. (\ref{aAsymp})-(\ref{BAsymp}).
The powers $s$, $p$ and $q$ are not shown since they are given in an explicit form.  
Further properties of the solutions are presented on Figs. \ref{data_mono_bis} and \ref{alpha_vary}. 
More specifically,
these figures summarize the dependence of the asymptotic coefficients on $\lambda$ and on $\alpha$.

\section{Conclusion}
\setcounter{equation}{0}

We have analyzed several types of spherically symmetric solutions of the SU(2) gauge theory coupled to
Conformal Gravity: pure YM solutions and monopole-like solutions in the YMH system.

To our knowledge, the pure YM solutions have not been discussed previously in the literature. These solutions are well 
localized and have a well-defined ``inertial mass'' as well as finite coefficients $c_1$ and $c_2$ of the 
``exterior solution''. Solutions exist for all values of $\alpha$ including 0 and negative ones, and for all finite $a_0$. 
Their magnetic charges are therefore continuous. These solutions contain also the Abelian purely magnetic
solutions \cite{mannheim_kazanas_2} in a singular limit. 

The monopole-like solutions exhibit a much longer range behavior due to the scalar field. As a result, their 
gravitational fields do not approach the vacuum solution (\ref{ConfSch}) and they do not posses a finite inertial mass. 
They exist for all values of $\alpha$ in the range $-4\lambda <\alpha < 3\lambda/2$. 
 These solutions are the closest possible analogues to the self-gravitating monopoles in 
the GR-YMH system. The significant differences which still exist seem to emerge mainly from the fact that the mechanism for 
symmetry breaking relies heavily on the presence of the non-minimal coupling to gravity.

The kind of equations we have solved (fourth-order) is  unconventional but could be treated with a good accuracy
by our numerical methods which in this case are indispensable.

A natural extension of this work would be to construct
  the black hole counterpart of the solutions, i.e.
  solutions with  the metric function $B(r)$ 
  presenting  a regular horizon at $r=r_h$; that is  $B(r_h)=0$.
  In the case of monopole, it would be interesting to study the
  dependence of the extremal values of $\alpha$ on the horizon size
  and to see if the domain of existence of conformal monopole-black-hole 
  fits in a pattern similar to the one of Fig.7  of Ref. \cite{BFM1992}.
     
\vspace{0.2cm}

\noindent {\large\textbf{acknowledgment}}

{\noindent One of us (Y.B.) thanks the Belgian FNRS for financial support.}


\begin{thebibliography}{8}

\bibitem{Mannheim2006}
  P. Mannheim,
 Prog.\ Part.\ Nucl.\ Phys.  {\bf 56}, 340 (2006).\vspace{-0.2cm}
 
 \bibitem{BrihayeVerbin}
  Y.~Brihaye and Y.~Verbin,
   Phys.\ Rev.\  D {\bf 80}, 124048 (2009). \vspace{-0.2cm}

\bibitem{MannKaz1989}
  P. Mannheim and D. Kazanas,
  Astrophys.\ J.\  {\bf 342}, 635 (1989).\vspace{-0.2cm}

\bibitem{Schmidt2006}
  H.~J.~Schmidt,
 Int.\ J.\ Geom.\ Meth.\ Mod.\ Phys.\  {\bf 4}, 209 (2007). \vspace{-0.2cm}

\bibitem{Fabbri2008}
  L.~Fabbri,
 {\em Higher-Order Theories of Gravitation },
  arXiv:0806.2610 [hep-th]. \vspace{-0.2cm}

\bibitem{BMK}
  R.~Bartnik and J.~McKinnon,
  Phys.\ Rev.\ Lett.\  {\bf 61}, 141 (1988). \vspace{-0.2cm}

\bibitem{bjoraker}
  J.~Bjoraker and Y.~Hosotani,
  Phys.\ Rev.\ Lett.\  {\bf 84}, 1853 (2000). \vspace{-0.2cm}

\bibitem{mannheim_kazanas_2}  
P. Mannheim and D. Kazanas, 
Phys.\ Rev.\  D {\bf 44}, 417 (1991). \vspace{-0.2cm}
  
  \bibitem{Ortiz1992}
  M.~E.~Ortiz,
  Phys.\ Rev.\  D {\bf 45}, R2586 (1992). \vspace{-0.2cm}
  
  \bibitem{LNW1992}
  K.~M.~Lee, V.~P.~Nair and E.~J.~Weinberg,
  Phys.\ Rev.\  D {\bf 45}, 2751 (1992). \vspace{-0.2cm}
  
 \bibitem{BFM1992}
  P.~Breitenlohner, P.~Forgacs and D.~Maison,
  Nucl.\ Phys.\  B {\bf 383}, 357 (1992). \vspace{-0.2cm}
  
\bibitem{BFM1995}
  P.~Breitenlohner, P.~Forgacs and D.~Maison,
  Nucl.\ Phys.\  B {\bf 442}, 126 (1995). \vspace{-0.2cm}
  
  
\bibitem{LugoEtAl1}
  A.~R.~Lugo and F.~A.~Schaposnik,
  Phys.\ Lett.\  B {\bf 467}, 43 (1999). \vspace{-0.2cm}
  
\bibitem{LugoEtAl2}
  A.~R.~Lugo, E.~F.~Moreno and F.~A.~Schaposnik,
  Phys.\ Lett.\  B {\bf 473}, 35 (2000). \vspace{-0.2cm}

\bibitem{EderyEtAl2006}
  A.~Edery, L.~Fabbri and M.~B.~Paranjape,
  Class.\ Quant.\ Grav.\  {\bf 23}, 6409 (2006). \vspace{-0.2cm}
   
 \bibitem{EderyEtAl2009}
  A.~Edery, L.~Fabbri and M.~B.~Paranjape,
  Can.\ J.\ Phys.\  {\bf 87}, 251 (2009). \vspace{-0.2cm}
  
 
\end{thebibliography}
\end{document}